\documentclass[final,3p,times,twocolumn]{elsarticle}
 \biboptions{comma,sort&compress}
\usepackage{here}
 \usepackage{graphicx}
  \usepackage{epsfig}

\def\beq{\begin{equation}}
\def\eeq{\end{equation}}
\def\bea{\begin{eqnarray}}
\def\eea{\end{eqnarray}}

\journal{Nuc. Phys. (Proc. Suppl.)}

\newcommand{\beqn}{\begin{eqnarray}}
\newcommand{\eeqn}{\end{eqnarray}}
\newcommand{\be}{\begin{equation}}
\newcommand{\ee}{\end{equation}}
\newcommand{\eqn}[1]{(\ref{#1})}

\newcommand{\ba}{\begin{array}{c}}
\newcommand{\bat}{\begin{array}{cc}}
\newcommand{\ea}{\end{array}}

\newcommand{\bi}{\begin{itemize}}
\newcommand{\ei}{\end{itemize}}
\newcommand{\chpt}{$\chi$PT}
\newcommand{\rcht}{R$\chi$T}

\newcommand{\cO}{{\cal O}}

\newcommand{\ket}{\,\rangle}
\newcommand{\bra}{\langle \,}

\begin{document}

\begin{frontmatter}



\title{Revisiting the vector form factor at next-to-leading order in $1/N_C$ \tnoteref{conference}}

\tnotetext[conference]{Talk given at the 15th International QCD Conference, 28th June --3d July (2010), Montpellier (France). IFIC/10-31 report.}
 \author{Ignasi Rosell}
 
 \address{Departamento de Ciencias F\'\i sicas, Matem\'aticas y de la Computaci\'on, 
Universidad CEU Cardenal Herrera, \\ c/ Sant Bartomeu 55, 
E-46115 Alfara del Patriarca, Val\`encia, Spain }

\address{IFIC, Universitat de Val\`encia - CSIC,
Apt. Correus 22085, E-46071 Val\`encia, Spain}

\ead{rosell@uch.ceu.es}

\begin{abstract}
Using the Resonance Chiral Theory lagrangian, we perform a calculation of the vector form factor of the pion at the next-to-leading order (NLO) in the $1/N_C$ expansion. Imposing the correct QCD short-distance constraints, one determines it in terms of $F$, $G_V$, $F_A$ and resonance masses. Its low momentum expansion fixes then the low-energy chiral couplings $L_{9}$ and $C_{88}-C_{90}$ at NLO, keeping full control of their renormalization scale dependence. At $\mu_0=0.77$~GeV, we obtain $L_{9}^r(\mu_0) = (7.6 \pm 0.6)\cdot 10^{-3}$ and $C_{88}^r(\mu_0)-C_{90}^r(\mu_0)=(-4.5 \pm 0.5)\cdot 10^{-5}$.
\end{abstract}


\end{frontmatter}



\section{Introduction}

Effective field theory is nowadays the standard technique to investigate the low-energy dynamics of QCD. In particular, the chiral symmetry constraints encoded in Chiral Perturbation Theory (\chpt) provide a very powerful tool
to access the non-perturbative regime through a perturbative expansion in powers of light quark masses and momenta \cite{Weinberg,ChPTp4}. The precision required in present phenomenological applications makes necessary to include corrections of $\cO(p^6)$~\cite{ChPTp6}. However,
the large number of unknown low-energy couplings (LECs) appearing at this order puts a clear limit to the achievable accuracy. A dynamical determination of these \chpt\ couplings is compulsory to achieve further progress in our understanding of strong interactions at low energies.

A useful connection between \chpt\ and the underlying QCD dynamics can be established in the limit of an infinite number of quark colours \cite{polychromatic,MHA}. Assuming confinement, the strong dynamics at $N_C\to\infty$ is given by tree diagrams with infinite sums of hadron exchanges, which correspond to the tree approximation of some local phenomenological lagrangian \cite{NC}. Resonance Chiral Theory (\rcht) provides a correct framework to incorporate these massive mesonic states within a phenomenological lagrangian formalism \cite{RChTa,RChTc}. Integrating out the heavy fields one recovers at low energies the \chpt\ lagrangian with explicit values of the chiral LECs in terms of resonance parameters. Since the short-distance properties of QCD impose stringent constraints on the \rcht\ couplings, it is then possible to extract information on the low-energy \chpt\ parameters.

Clearly, we cannot determine the infinite number of meson couplings which characterize the large--$N_C$ lagrangian. 
However, one can obtain useful approximations in terms of a finite number of meson fields. Truncating the infinite tower of meson resonances to the lowest states, 
one gets a very successful prediction of the $\cO(p^4 N_C)$ \chpt\ couplings in terms of a few resonance parameters \cite{polychromatic}. 

Since chiral loop corrections are of next-to-leading order (NLO) in the $1/N_C$ expansion, the large--$N_C$ determination of the LECs is unable to control their renormalization-scale dependence. 
First analyses of resonance loop contributions to the running of $L_{10}^r(\mu)$ and $L_9^r(\mu)$ were attempted in ref.~\cite{CP:02} and ref.~\cite{RSP:05} respectively. In spite of all the complexity associated with the still not so well understood renormalization of \rcht\ \cite{RSP:05,RPP:05,saturation,natxo-tesis}, these pionnering calculations have shown the potential
predictability at the NLO in $1/N_C$.

Using dispersion relations we can avoid technicalities associated with the renormalization procedure. 
Following these ideas, in \cite{L8L10-nlo} we determined, respectively, the couplings $L_8^r(\mu)$, $C_{38}^r(\mu)$ and $L_{10}^r(\mu)$, $C_{87}^r(\mu)$ at NLO in $1/N_C$. As a next step, we present here the study of the vector form factor of the pion, which allows to perform a NLO determination of the related chiral couplings.


\section{The lagrangian}

Let us adopt the Single Resonance Approximation, where just the lightest resonances are considered.
On account of large-$N_C$, the mesons are put together into $U(3)$ multiplets. Hence, our degrees of freedom are the pseudo-Goldstone bosons 
along with massive multiplets of the type $V(1^{--})$, $A(1^{++})$, $S(0^{++})$ and $P(0^{-+})$. With them, we construct the most general action that preserves chiral symmetry. Since we are interested on the structure of the interaction at short distances, we will work in the chiral limit. With this simplification we do not loose any information on the LECs we want to determine, because they are independent of the light quark masses.

The Resonance Chiral Theory must satisfy the high-energy behaviour dictated by QCD. To comply with this requirement we will only consider operators constructed with chiral tensors of $\cO(p^2)$; interactions with higher-order chiral tensors would violate the QCD asymptotic behaviour, unless their couplings are severely fine tuned to ensure the needed cancellations at high energies \cite{saturation}. 

The different terms in the lagrangian can be classified by their number of resonance fields:
\begin{eqnarray} \label{lagrangian}
\mathcal{L}_{R\chi T}&=&\mathcal{L}_\chi \,+\,\sum_{R_1}\mathcal{L}_{R_1}
\,+\,\sum_{R_1,R_2}\mathcal{L}_{R_1R_2}
\, + \, ... \,\,\,  ,
\end{eqnarray}
where the dots denote the irrelevant operators with three or more resonance fields, and the indices $R_i$ run over all different resonance multiplets, $V$, $A$, $S$ and $P$. The $\cO(p^2)$ $\chi$PT lagrangian~\cite{ChPTp4},
\begin{eqnarray}
\mathcal{L}_{\chi} &=& \frac{F^2}{4} \bra u_\mu u^\mu + \chi_+ \ket  \,,
\end{eqnarray}
contains the terms with no resonance fields. The second term in \eqn{lagrangian} corresponds to the operators with one massive resonance~\cite{RChTa},
\begin{eqnarray}
\mathcal{L}_V &=& \frac{F_V}{2\sqrt{2}} \bra V_{\mu\nu} f^{\mu\nu}_+ \ket + \frac{i\, G_V}{2\sqrt{2}} \bra V_{\mu\nu} [u^\mu, u^\nu] \ket  \, , \nonumber \\
\mathcal{L}_A &=& \frac{F_A}{2\sqrt{2}} \bra A_{\mu\nu} f^{\mu\nu}_- \ket \, ,\nonumber \\
\mathcal{L}_S &=& c_d \bra S u_\mu u^\mu\ket + c_m\bra S\chi_+\ket \,, \phantom{\frac{1}{2}} \nonumber  \\
\mathcal{L}_P &=& i\,d_m \bra P \chi_- \ket \,.\phantom{\frac{1}{2}} \label{P}
\end{eqnarray}
The lagrangian $\mathcal{L}_{R_1R_2}$ contains the kinetic resonance terms and the remaining operators with
two resonance fields~\cite{RChTa,RChTc,RSP:05}. We show only the relevant operators for the vector form factor of the pion, taking into account that here we just consider the lowest-mass two-particle exchanges: two Goldstone bosons or one Goldstone and one resonance field (exchanges of two heavy resonances are kinematically suppressed),
\begin{eqnarray}
 \mathcal{L}_{SA}\!\!&=\!\!& \lambda^{SA}_1 \bra \{\nabla_\mu S, A^{\mu\nu} \} u_\nu \ket   \, , \nonumber \\
 \mathcal{L}_{PV}\!\!&=\!\!& i \lambda^{PV}_1\bra [\nabla^\mu P,V_{\mu\nu} ] u^\nu \ket   \, ,  \nonumber \\
 \mathcal{L}_{VA}\!\!&=\!\!&  i \lambda^{VA}_2 \! \bra \!  [ V^{\mu\nu}, A_{\nu\alpha} ] h^\alpha_\mu \!\ket  \! +\! i \lambda^{VA}_3 \! \bra \! [ \nabla^\mu V_{\mu\nu}, A^{\nu\alpha} ] u_\alpha \!\ket   \nonumber  \\
&&+ i \lambda^{VA}_4 \bra   [ \nabla_\alpha V_{\mu\nu}, A^{\alpha\nu} ] u^\mu \ket   \nonumber \\ && + i \lambda^{VA}_5 \bra  [ \nabla_\alpha V_{\mu\nu}, A^{\mu\nu} ] u^\alpha \ket \,.  \end{eqnarray}
%
The brackets $\langle ... \rangle$ denote a trace of the corresponding flavour matrices and the standard notation defined in refs.~\cite{RChTa,RChTc} is adopted. Note that
our lagrangian $\mathcal{L}_{R\chi T}$ satisfies the $N_C$ counting rules for
a theory with $U(3)$ multiplets, so that only operators that have one trace
in the flavour space are considered. 


The \rcht\ lagrangian~\eqn{lagrangian} contains a large number of unknown coupling constants. However, as we will see in the next section, the short-distance QCD constraints allow to determine many of them. In the observable at hand and with our assumptions, we have nine couplings or combinations of them ($F$, $F_V$, $G_V$, $F_A$, $c_d$, $\lambda^{SA}_1$, $\lambda^{PV}_1$, $-2 \lambda_2^{\mathrm{VA}} + \lambda_3^{\mathrm{VA}}$ and $2 \lambda_2^{\mathrm{VA}} -2 \lambda_3^{\mathrm{VA}} + \lambda_4^{\mathrm{VA}} + 2 \lambda_5^{\mathrm{VA}}$) and four resonance masses ($M_V$, $M_A$, $M_S$ and $M_P$). As we will see in section~\ref{sec:VFF}, after imposing a good short-distance behaviour, the number of parameters has been reduced to three couplings  ($F$, $G_V$ and $F_A$) and three masses ($M_V$, $M_A$ and $M_S$). The role of the information coming from the underlying theory is fundamental.

\section{The vector form factor of the pion} \label{sec:VFF}

Our observable is defined through the two pseudo-Goldstone matrix element of the vector current:
\begin{equation}
\bra  \pi^+(p_1)\,  \pi^-(p_2)\, |\, J^\mu_V \,|\, 0  \ket   =  \mathcal{F}(s)\, (p_1-p_2)^\mu \, , 
\end{equation}
where $J^\mu_V = \frac{1}{2}\left( \bar{u}\gamma^\mu u - \bar{d}\gamma^\mu d \right)$ and $s\equiv (p_1+p_2)^2$. At very low energies, $\mathcal{F}(s)$ has been studied within the $\chi$PT framework up to $\cO(p^6)$~\cite{ChPTp4,VFF_ChPT}. A first study at NLO within R$\chi$T and the $1/N_C$ expansion  was done in ref.~\cite{RSP:05}.

It is important to remark the convenience of improving the vector form factor at subleading order at intermediate energies. In ref.~\cite{RSP:05} only operators with up to one resonance field were including, so that $\mathcal{F}(s)$ was bad-behaved at high energies, since it was not possible to get a proper form factor at high energies without the inclusion of new operators. Furthermore,
in the final result of \cite{RSP:05} there were free couplings, the equivalent resonance couplings to the $L_9$ and $(C_{88}-C_{90})$ LECs. Now we know that we do not need to consider them as soon as the observable is well behaved at short distances~\cite{saturation}.

Within Resonance Chiral Theory and at leading-order in $1/N_C$, one finds~\cite{RChTa}
\begin{equation}
\mathcal{F}_{R \chi T} (s) =  1 + \frac{F_VG_V}{F^2} \frac{s}{M_V^2 - s} \,.
\end{equation}
Considering that the form factor is constrained to be zero at infinite momentum transfer~\cite{brodsky-lepage}, the vector couplings should satisfy $F_V G_V = F^2$,
which implies~\cite{RChTa}
\begin{equation}
\mathcal{F}_{R \chi T} (s) =  \frac{M_V^2}{M_V^2 - s} \,. \label{VFFLO}
\end{equation}

The subleading corrections can be calculated by means of dispersive relations. Once the NLO absorptive parts of $\mathcal{F}_{R \chi T} \left(s\right)$ are known, one can reconstruct the full form factor up to appropriate subtraction terms,
\begin{equation}
\mathcal{F}_{R \chi T} \left(s\right) =  \frac{M_V^2}{M_V^2 - s}  + \Delta  \mathcal{F}(s)  \,, \label{VFFRChT}
\end{equation}
being $ \Delta \mathcal{F}(s)$ the NLO contribution coming from the usual dispersive relation after imposing a good behaviour at short distances,
\begin{equation}
\Delta \mathcal{F}(s) = \frac{s}{M_V^2 - s} \,\delta_{\mathrm{NLO}}   \,+\,  \frac{s}{\pi} \int_0^{\infty} \mathrm{d} t \,\frac{\mathrm{Im} \,\mathcal{F} (t)  }{t(t-s)} \,,  \label{VFFRChTcuts}
\end{equation}
where really the singular part of the integral at $t\to M_V^2$ has been subtracted. $\delta_{\mathrm{NLO}}$ is fixed by imposing that $ \Delta \mathcal{F}(s)$ vanish at high energies. Note that the relevant subtraction constant $c$, of next-to-leading order,  has been reabsorbed into the tree-level contributions $\delta_{\mathrm{NLO}}$,
\begin{eqnarray}
\frac{F_V^r\,G_V^r}{F^2}+c &=&1 \,+\,  \delta_{\mathrm{NLO}} \,. \label{deltaNLO}
\end{eqnarray}
Then the procedure we have followed to calculate the vector form factor at NLO has four steps:
\begin{enumerate}
\item Determine the spectral function of the considered absorptive cuts. 
\item Constraint the imaginary part to vanish at high energies. We have imposed vanishing spectral functions channel by channel.
In the case of the $\pi\pi$ cut we have found two constraints,
\begin{equation}
F_V G_V\,=\, F^2 \,, \qquad     3\,G_V^2+2\,c_d^2 \,=\, F^2 \,, \label{constraintpipi}
\end{equation}
where the first one coincides with the LO constraint. The second one is consistent with \cite{JJGuo}, obtained in the context of the $\pi\pi$ scattering at LO. 
In the case of the $P\pi$ cut the only possible solution is killing all the contribution by means of $\lambda_1^{\mathrm{PV}}$,
\begin{eqnarray}
\lambda_1^{\mathrm{PV}}\,=\,0\,. \label{constraintPpi}
\end{eqnarray}
For the $A\pi$ cut there is no only one solution, and we have used the solutions consistent with \cite{natxo-tesis,L8L10-nlo}, where tree-level form factors to resonance fields as asymptotic states where studied in the context of the correlator $\Pi(s)=\Pi_{VV}(s)-\Pi_{AA}(s)$ at NLO,
\begin{eqnarray}
-2 \lambda_2^{\mathrm{VA}} + \lambda_3^{\mathrm{VA}}&=& 0 \,, \phantom{\frac{1}{2}} \nonumber \\
- \lambda_3^{\mathrm{VA}} + \lambda_4^{\mathrm{VA}} + 2 \lambda_5^{\mathrm{VA}} &=& \frac{F_A}{F_V} \,, \nonumber \\
-\frac{F_A \,G_V\left(M_A^2 - 4\,M_V^2\right)}{3\sqrt{2} M_A^2 c_d\,F_V}&=& \lambda_1^{\mathrm{SA}} \,. \label{constraintApi}
\end{eqnarray}
Note that after considering the relations (\ref{constraintpipi}), (\ref{constraintPpi}) and (\ref{constraintApi}) the spectral functions can be expressed in terms of $G_V$, $F_A$, $F$ and masses.
\item Now the spectral function is prepared for the dispersive relation, which allows to reconstruct the full form factor up to $\delta_{\mathrm{NLO}}$ in eq.~(\ref{VFFRChTcuts}).
\item Finally, we impose that $\mathcal{F}_{R \chi T} (s)$ vanishes at short distances, what fixes $\delta_{\mathrm{NLO}}$ in (\ref{deltaNLO}).
\end{enumerate}

\section{The chiral couplings $L_{9}^r(\mu)$ and $C_{88}^r(\mu)-C_{90}^r(\mu)$} \label{sec:L9}

The low-momentum expansion of $\mathcal{F}(s)$ is determined by $\chi$PT~\cite{ChPTp4,VFF_ChPT}:
\begin{eqnarray}
\mathcal{F}_{\chi PT}  \!\!&=\!\!&  1 + \displaystyle\frac{2\,s}{F^2} \left\{L_9^r +\frac{\Gamma_9}{32\pi^2}  \left( \frac{5}{3}-\log \frac{-s}{\mu^2} \right) \right\} \nonumber \\&& -\displaystyle\frac{4\,s^2}{F^4} \left\{ C_{88}^r - C_{90}^r -\frac{\Gamma_{88}^{(L)}-\Gamma_{90}^{(L)} }{32\pi^2} \times \right. \nonumber \\ &&\quad \left.
\left( \frac{5}{3}-\log \frac{-s}{\mu^2} \right) +\mathcal{O}\!\left(N_C^{0}\right) \right\} +
 \mathcal{O}\! \left(s^3\right)\, ,  \nonumber \\
\label{VFFChPT}
\end{eqnarray}
with $\Gamma_{9} = 1/4$~\cite{ChPTp4} and  $\Gamma_{88}^{(L)} -\Gamma_{90}^{(L)}= -2L_1/3+L_2/3-L_3/2+L_9/4$~\cite{ChPTp6}. 



At LO in $1/N_C$ and within $\chi$PT, eq.~(\ref{VFFChPT}) becomes
\begin{equation}
\mathcal{F}_{\chi PT}  = 1 + \frac{2\,s}{F^2}  L_9  - \frac{4\,s^2}{F^4}   \left( C_{88} - C_{90} \right) + \mathcal{O} \!\left(s^3\right) . \label{ChPTLO}
\end{equation} 
Also at leading-order and now within R$\chi$T, eq.~(\ref{VFFLO}) can be expanded at low energies,
\begin{equation}
\mathcal{F}_{R \chi T}  = 1 + \frac{s}{M_V^2} + \frac{s^2}{M_V^4} + \mathcal{O}\! \left(s^3\right)   \,. \label{expansionLO}
\end{equation}
The matching between (\ref{ChPTLO}) and (\ref{expansionLO}) fixes $L_9$ and $C_{88}-C_{90}$ at LO~\cite{RChTa,RChTc},
\begin{equation}
L_9\,=\,\frac{F^2}{2M_V^2} \,,  \qquad C_{88}-C_{90}\,=\, -\frac{F^4}{4M_V^4} \,. \label{L10LO}
\end{equation}
Using $M_V\simeq 0.77\,$GeV and $F\simeq 89\,$MeV, one gets the large-$N_C$ estimates: $L_9\simeq 6.7 \cdot 10^{-3} $ and $C_{88}-C_{90} \simeq -4.5 \cdot 10^{-5}$.  At $\mu_0=770$~MeV,  the phenomenological determinations $L_9^r(\mu_0)=\left( 6.9\pm 0.7 \right) \cdot 10^{-3}$~\cite{ChPTp4,polychromatic} and $L_9^r(\mu_0)=\left(5.93 \pm 0.43\right)\cdot 10^{-3}, C_{88}^r (\mu_0)-C_{90}^r (\mu_0) =\left( -5.5 \pm 0.5 \right) \cdot 10^{-5}$~\cite{VFF_ChPT}, obtained respectively from a $\mathcal{O}(p^4)$ and a $\mathcal{O}(p^6)$ ChPT fit, agree approximately with the LO estimates.


Following the same steps as before, let us determine the related low-energy constants by matching eq.~(\ref{VFFChPT}) and the low-energy expansion of eq.~(\ref{VFFRChT}),
\begin{eqnarray}
\mathcal{F}_{R \chi T} \!\!\!&=\!\!\!& 1 \!+\! \displaystyle\frac{2s}{F^2} \! \left\{ \frac{F^2}{2M_V^2} \!+\!  \xi^{(2)} \!+\!\frac{\Gamma_9}{32\pi^2} \!\left( \frac{5}{3}\!-\!\log \frac{-s}{M_V^2} \right)  \right\} \nonumber \\ &&   - \displaystyle\frac{4\,s^2}{F^4} \left\{  -\frac{F^4}{4M_V^4} +   \xi^{(4)} -\frac{\Gamma_{88}^{(L)}-\Gamma_{90}^{(L)} }{32\pi^2} \times \right. \nonumber \\ && \qquad \quad \left. 
\left( \frac{5}{3}-\log \frac{-s}{M_V^2} \right) \right\} +
\mathcal{O}\! \left(s^3\right)\, , 
\label{VFFRChTexpansion}
\end{eqnarray}
where $\xi^{(n)}$ are the relevant $\mathcal{O}(s^n)$ coefficient of the low-energy expansion of $\Delta \mathcal{F}(s)$, once the structure coming from the $\chi$PT one-loop diagram has been subtracted. Now it is straightforward to estimate the LECs: 
\begin{eqnarray}
L_{9}^r(M_V^2) &=& \frac{F^2}{2M_V^2}+   \xi^{(2)} \,, \nonumber \\
C_{88}^r(M_V^2)-C_{90}^r(M_V^2) &=& -\frac{F^4}{4M_V^2}+ \xi^{(4)} \,.
\label{L9NLO}
\end{eqnarray}
 
 \begin{table}
\begin{center}
\begin{tabular}{|c|c|c|}
\hline
  &  $10^3 \cdot L_9^r (\mu_0 ) $  & $10^5 \cdot \left( C_{88}^r (\mu_0) - C_{90}^r (\mu_0) \right)$    \\[5pt]
\hline
This work &$7.6 \pm 0.6 $ & $-4.5 \pm 0.5$ \\
ref.~\cite{ChPTp4} & $6.9 \pm 0.7 $ & \\
ref.~\cite{VFF_ChPT}  & $5.93 \pm 0.43 $ &  $-5.5 \pm 0.5 $ \\
ref.~\cite{preVFF2} & $7.04 \pm 0.23 $ & \\
ref.~\cite{GA} & $6.54 \pm 0.15$ & \\
 & $5.50 \pm 0.40 $ & \\ 
 %
\hline
\end{tabular}
\end{center}
\caption{Comparison of different determinations, with $\mu_0=770\,$MeV.}
\label{tab:tab2}
\end{table}
 
We take the ranges~\cite{ChPTp4,PDG} $M_V=\left( 770 \pm 5 \right)\,$MeV, $M_S=\left(1090 \pm 110 \right)\,$MeV, $M_A= \left(1200 \pm 200 \right)\,$MeV  and $F=\left(89\pm2 \right)\,$MeV. From the observed rate $\Gamma\left( a_1 \to \pi \gamma \right)= (650 \pm 250 )\,$keV~\cite{PDG}, one is able to estimate the vector-axial coupling, $F_A=(120\pm 20 )\,$MeV. The constraint of (\ref{constraintpipi}) implies that $G_V < F/\sqrt{3}$, so that we take the range $G_V \in [40,50]\,$MeV. This gives the numerical prediction:
\begin{eqnarray}
L_9^r (\mu_0 ) &=& \left( 7.6 \pm 0.6 \right) \cdot 10^{-3} \, , \nonumber \\
 C_{88}^r (\mu_0) - C_{90}^r (\mu_0)  &=& \left( -4.5 \pm 0.5 \right) \cdot 10^{-5} \,, \label{pheno1}
\end{eqnarray}
being $\mu_0$ the usual renormalization scale, $\mu_0=770\,$MeV.

\section*{Acknowledgements}

I wish to thank S.~Narison for the organization of the conference and A.~Pich and J.J.~Sanz-Cillero for their helpful comments. This work has been supported by the Universidad CEU Cardenal Herrera, by the Spanish Government (FPA2007-60323 and Consolider-Ingenio 2010 CSD2007-00042, CPAN), and by the European Union (MRTN-CT-2006-035482, FLAVIAnet).












\end{document}